\documentclass[aps,prl,notitlepage,twocolumn]{revtex4-1}
\usepackage{graphicx}
\usepackage[colorlinks=true, allcolors=magenta]{hyperref}
\usepackage{xcolor}
\usepackage{amsmath,amssymb,bm}
\usepackage[bbgreekl]{mathbbol}
\usepackage{textgreek}
\usepackage[utf8]{inputenc}
\newcommand{\sectionn}[1]{\textit{#1}}

\usepackage{physics}
\usepackage{mathrsfs}
\usepackage{comment}

\renewcommand{\vr}{\mathbf{r}}
\newcommand{\vk}{\mathbf{k}}
\newcommand{\vR}{\mathbf{R}}
\newcommand{\vK}{\mathbf{K}}
\newcommand{\vB}{\mathbf{B}}
\newcommand{\tB}{\mathcal{B}}
\newcommand{\nbl}{\bm{\nabla}}

\renewcommand{\(}{\begin{equation}\begin{aligned}}
\renewcommand{\)}{\end{aligned}\end{equation}}

\begin{document} 

\title{{Flavors of Magnetic Noise in Quantum Materials}}
\author{Shu Zhang}
\email{suzy@physics.ucla.edu}
\affiliation{Department of Physics and Astronomy, University of California, Los Angeles, California 90095, USA}
\author{Yaroslav Tserkovnyak}
\affiliation{Department of Physics and Astronomy, University of California, Los Angeles, California 90095, USA}
\date{\today}

\begin{abstract}
    {The complexity of electronic band structures in quantum materials offers new charge-neutral degrees of freedom stable for transport, a promising example being the valley (axial) degree of freedom in Weyl semimetals (WSMs). 
    A noninvasive probe of their transport properties is possible by exploiting the frequency dependence of the magnetic noise generated in the vicinity of the material.
    In this work, we investigate the magnetic noise generically associated with diffusive transport using a systematic Langevin approach.
    Taking a minimal model of magnetic WSMs for demonstration,
    we show that thermal fluctuations of the charge current, the valley current, and the magnetic order can give rise to magnetic noise with distinctively different spectral characters, which provide a theoretical guidance to separate their contributions. 
    Our approach is extendable to the study of magnetic noise and its spectral features arising from other transport degrees of freedom in quantum materials.}
\end{abstract}

\maketitle
\sectionn{Introduction.}|{Many recently discovered novel quantum materials are featured by the complexity of their electronic band structures, as a result of spin-orbit coupling~\cite{hasan2010}, magnetic order~\cite{watanabe2018}, twist engineering~\cite{Zou2018}, etc. These structures may exhibit new degrees of freedom stable for transport, in addition to charge and spin, due to the protection either by symmetry or topology, which can be explored for the next-generation information devices. One prominent example is the valley degree of freedom present in some hexagonal two-dimensional semiconductors or Weyl semimetals (WSMs), the possibility of manipulating which gives rise to the field of valleytronics~\cite{valleytronics}.}

{Transport is generally noisy. As the Johnson-Nyquist noise (thermally excited electric currents) is important in electronic devices, understanding the generation of noise from these new degrees of freedom is of practical relevance in spintronic or valleytronic devices. At the same time,}
the electromagnetic noise emitted by a material into its environment encodes rich information about its intrinsic excitation dynamics and transport properties. 
For example, 
the magnetic noise in the vicinity of a conductor
is directly related to its impedance~\cite{Johnson1928,varpula1984}. 
The recent development of magnetic noise spectroscopy using single qubits, especially the nitrogen-vacancy (NV) centers in diamond~\cite{NV2008}, has provided a nanoscale probe to access such information noninvasively, and with high frequency resolution~\cite{JohnsonNV2015,Agarwal2017,Jayich2018}.
NVs 
have also turned out to be useful in the study of magnetic insulators~\cite{Yacoby2015,Du2017,Yacoby2018,Chatterjee2019} by probing the magnetic noise generated by spin excitations.
{While thermodynamic valley fluctuations have been accessed by optical methods~\cite{valley-noise2019}, the noise associated with valley transport in low-frequency regimes is so far rarely explored.}

\begin{figure}[t]
    \centering
    \includegraphics[width = \linewidth]{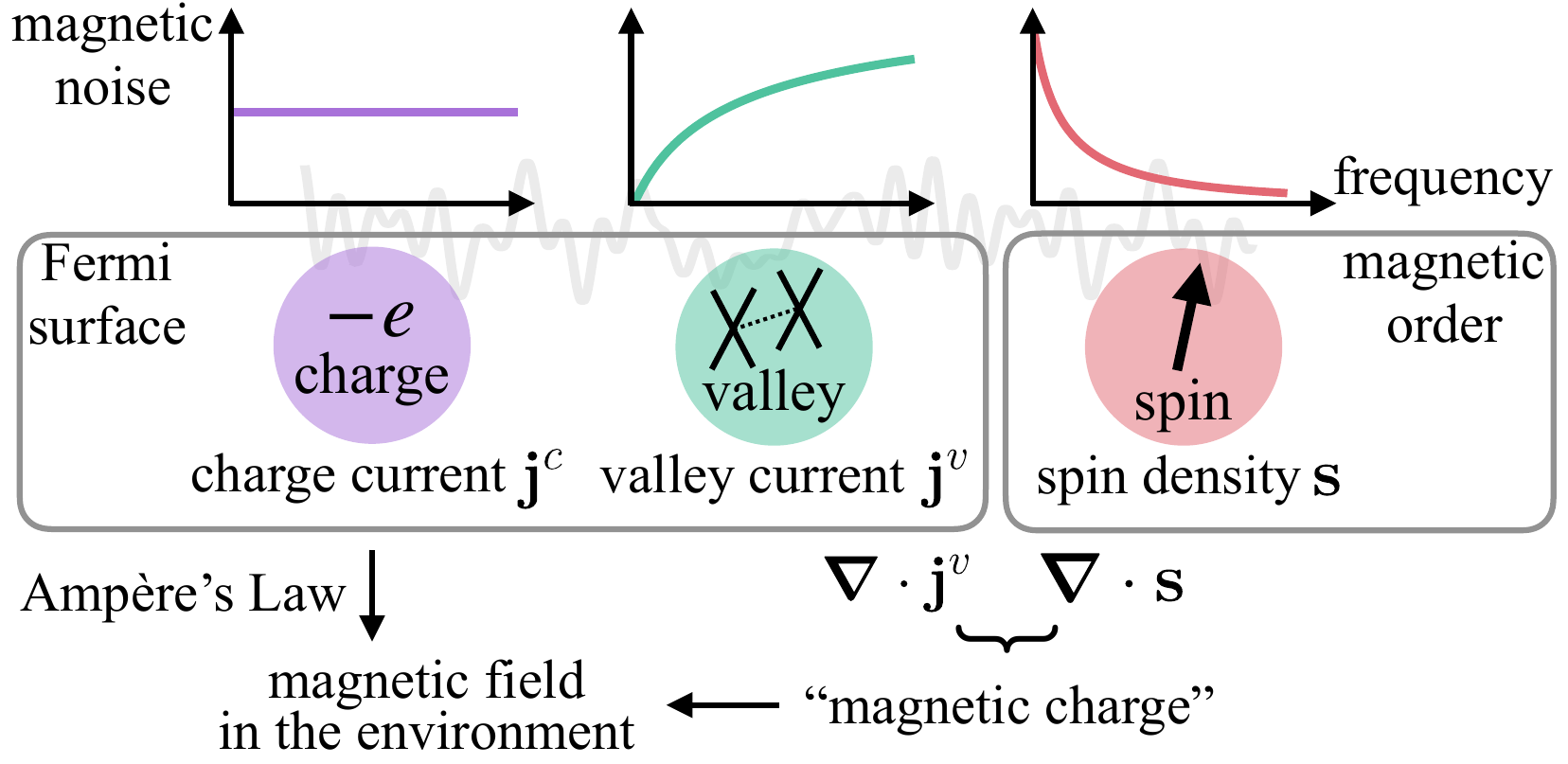}
    \caption{An illustration of the three flavors of magnetic noise in a magnetic WSM and their different spectral characters.}
    \label{fig:scheme}
\end{figure}

{In this work, we offer a qualitative perspective to the study of the magnetic noise in quantum materials, which can be flavorful due to the presence of various transport degrees of freedom, focusing on their generic diffusive aspect. To this end, we take an example of a magnetic WSM, which naturally involves three sources of noise, namely, charge, spin, and valley. Intriguingly, each flavor can contribute a distinct spectral character, as shown in Fig.~\ref{fig:scheme}. Our perspective can be extended to consider other pseudospin degrees of freedom in general and may inspire future experimental work in the NV probe of quantum materials in light of its advantage in frequency resolution in the GHz regime. } 


WSMs are a family of topological quantum materials promising for valleytronic applications,
because the band crossing at Weyl points is topologically protected, and the valley relaxation time can be very long in a clean system~\cite{HOSUR2013}.
Magnetic WSMs, such as those in magnetic Heusler compounds~\cite{Bernevig2016,Felser2016,Felser2018review}, allow the existence of a single pair of {energy-degenerate} Weyl valleys~\cite{Canfield2019}, and are thus ideal for the study of valley transport. 
The interplay between charge, valley, and spin~\cite{SSZhang2019,YT2021} also makes them attractive for spintronics.
The detection of valley transport often relies on the conversion from valley excitations to optic or electric signals, for instance, with the help of the chiral anomaly effect in a nonlocal geometry~\cite{parameswaran2014}. 
In magnetic WSMs, however, this is not totally unambiguous due to the presence of spin excitations, which usually have a long diffusion length as well.
The magnetic noise spectroscopy can serve as a direct probe of the intrinsic transport properties in the absence of external perturbations and electric contacts. 
As we will show, it is possible to distinguish the valley and spin contributions to the magnetic noise based on their spectral characteristics.

\sectionn{Main concepts.}|We first briefly summarize our conceptual understanding of why the three flavors contribute differently, even all under a diffusive treatment. 
Focusing on the scenario with a nonvanishing carrier density, where the electric charge density fluctuations are screened by Coulomb interactions, transverse fluctuations of the charge current dominate the charge channel. In contrast, longitudinal fluctuations of the valley current are important due to {its charge neutrality and hence the absence of screening}
of the axial charge density. In our model (see below), 
the {Weyl nodes are induced} 
by the broken time-reversal symmetry associated with the magnetic order, which
dictates the response of valley currents to magnetic fields and thus the generation of magnetic noise by valley fluctuations.
Consequently, only the longitudinal component of the valley current generates magnetic fields in the environment: The valley current $\mathbf{j}^v$ behaves as a magnetization, and determines a ``magnetic charge'' distribution in the bulk $\rho_M \propto -\nbl \cdot \mathbf{j}^v$, while that on the surface vanishes due to the boundary condition $\sigma_M \propto \mathbf{n} \cdot \mathbf{j}^v = 0$. In the spin channel, however, the spin density (rather than the spin current) plays such a role. See Fig.~\ref{fig:scheme} for comparison.


\begin{figure}[t]
    \centering
    \includegraphics[width = \linewidth]{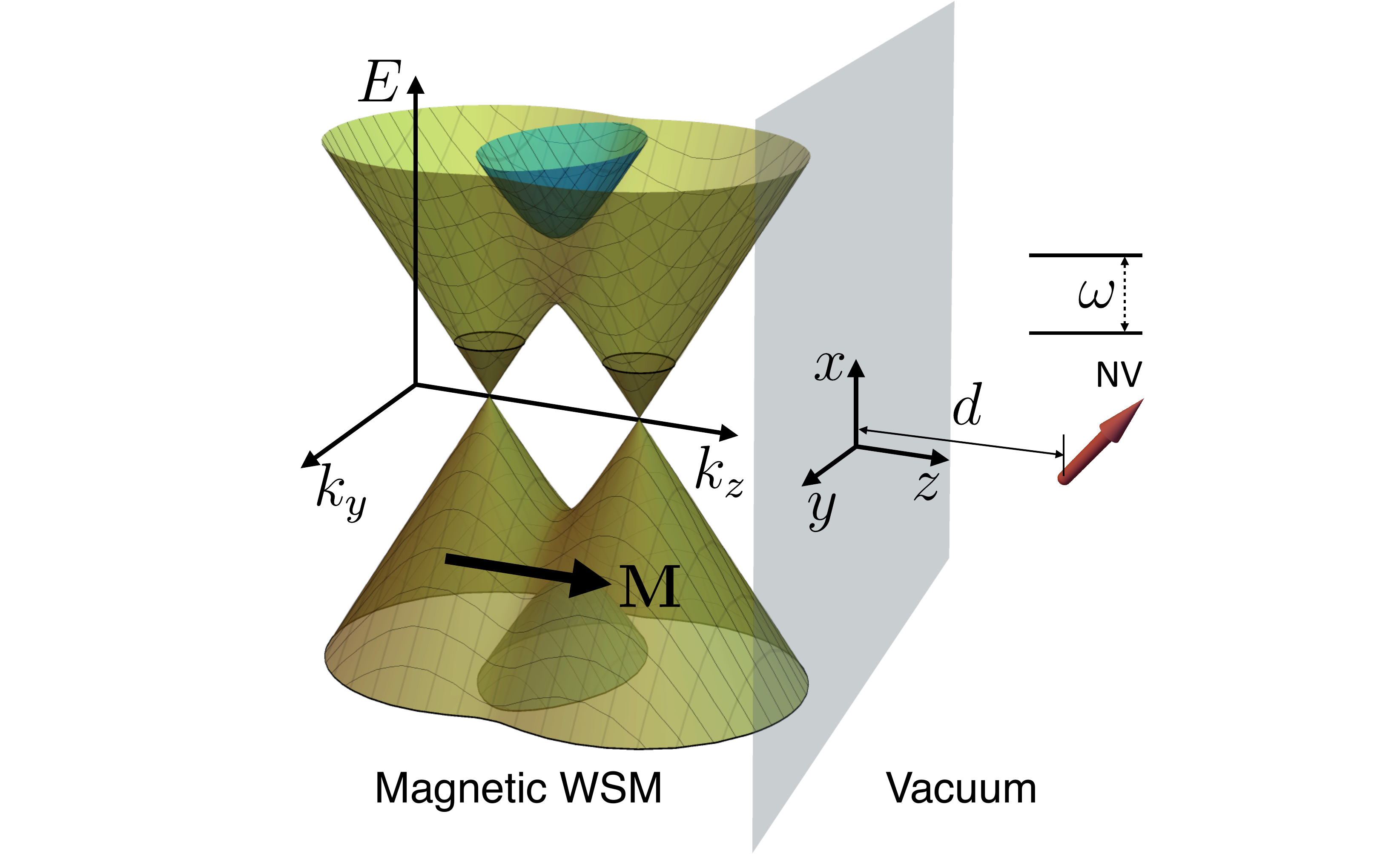}
    \caption{Quantum-impurity relaxometry of a magnetic Weyl semimetal. A three-dimensional sample of a magnetic WSM with a band structure given by Eq.~(\ref{eq:Hamiltonian}) has a magnetic order $\mathbf{M}$ in the $z$ direction and a surface lying in the $xy$ plane. An NV center is placed at distance $d$ from the surface, with a tunable resonance frequency $\omega$.}
    \label{fig:1}
\end{figure}

\sectionn{Model.}|The following minimal model is considered for a magnetic WSM with four bands~\cite{Burkov2011}:
\(
    H = v \tau_z \otimes (\bm{\sigma} \cdot \hbar \mathbf{k}) + \Delta \tau_x + J \bm{\sigma} \cdot \mathbf{M},
    \label{eq:Hamiltonian}
\)
where $\bm{\sigma}$ and $\bm{\tau}$ are vectors of the Pauli matrices in the spin and valley spaces, respectively, $v$ is the Fermi velocity, $\Delta$ is a Dirac mass, and $J$ is the magnetic exchange between the itinerant electrons and the magnetic order $\mathbf{M}$. 
When $|J|M > |\Delta|$, this model realizes a single pair of Weyl nodes with opposite chiralities separated in the momentum space. Here after, $|J|M \gg |\Delta|$ is assumed to approximately conserve the valley index~\cite{YT2021}. We also assume a small but finite Fermi surface, and thus a finite carrier density at the valleys, i.e., a Weyl metal.

Defining $\mathbf{j}^{+(-)}$ as the number flux density operator of electrons belonging to the valley with positive (negative) chirality, the valley current becomes
\(
    \mathbf{j}^v 
    = \mathbf{j}^+ - \mathbf{j}^-
    =  \tau_z \partial_\mathbf{k} H / \hbar 
    =  v \bm{\sigma},
    \label{eq:valley-current-spin}
\)
at a single-particle level.
The assumptions about the spin and parity symmetries~\cite{YT2021} in our model~(\ref{eq:Hamiltonian}), therefore, establish a proportionality between the valley current and the itinerant spin density at the Fermi surface. Since a Zeeman term $\bm{\sigma} \cdot \mathbf{B}$ is allowed, the valley current can directly couple to magnetic fields, with its fluctuations generating magnetic noise.

We employ a simple setup of magnetic noise measurement using an NV center placed at a nanoscale distance $d$ from a flat surface of the three-dimensional material, as shown in Fig.~\ref{fig:1}. 
We choose the magnetic order $\mathbf{M} \parallel \hat{\mathbf{z}}$ and the surface plane to be the $xy$ plane. With this geometry, we focus on the contributions from bulk transport, separated from the Fermi arcs. 
The central object of our study is the magnetic noise tensor in the frequency domain at the position of the NV center $\vr_\text{NV}$, expressed in the symmetrized correlation functions of the magnetic field operators,
\(
    \tB_{ii'} (\omega) 
    = \frac{1}{2} \int dt \, e^{i \omega t} 
    \langle \{B_i (\vr_\text{NV}, t), B_{i'}(\vr_\text{NV}, 0)\} \rangle.
    \label{eq:noise-correlation}
\)
Its components determine the NV relaxation rate depending on the NV orientation~\cite{Yacoby2015,Flebus2018}.
Keeping the typical GHz frequency of NV centers in mind, we focus on the magnetostatic limit, where the wavelength $\lambda$ of the fluctuating electromagnetic field and the skin depth $\lambda_s$ of the material are both much larger than $d$. 
This frequency also puts us in the classical limit ($\hbar \omega \ll k_B T$), with the exception only for very low temperatures.
Accordingly, we develop a simplified but systematic treatment of the magnetic-noise generation from charge, valley, and spin degrees of freedom, which we now turn to.

\sectionn{Charge.}|To introduce our approach, we first reproduce the established results for the Johnson-Nyquist noise. The Langevin dynamics~\cite{ChaikinLubensky1995,sinitsyn2016} of the electron current (the number flux density)
$\mathbf{j}^c = - \sigma \nbl \mu^c + \bm{\epsilon}^c$ is considered,
where $\sigma$ is the conductivity (neglecting the Hall effect~\footnote{A small Hall conductivity would not affect our result since magnetic noise is governed by dissipative properties. In the presence of a large Hall angle, the theory can be generalized using a full conductivity tensor.}), $\mu^c$ is the electrochemical potential, and $\bm{\epsilon}^c$ is a Gaussian white noise with 
$\langle \epsilon^c_i (\vr, t) \rangle = 0$ and 
$\langle \epsilon^c_i (\vr, t) \epsilon^c_{i'} (\vr', t') \rangle = 2 \sigma k_B T \delta_{ii'} \delta (\vr - \vr') \delta(t-t')$. The coefficient in the correlation follows from the equipartition theorem~\cite{Nyquist1928}, and is consistent with the fluctuation dissipation theorem~\cite{kubo1966}.
This treatment assumes the diffusive regime, {focusing on dynamics on the length scale} much larger than the mean free path of the electrons, and at frequencies much lower than the momentum relaxation rate.
Focusing on the transverse components of the charge current, $\nbl \cdot \mathbf{j}^c = 0$, we have the following differential equation for the electrochemical potential $\mu^c$:
\(
    \sigma \nbl^2 \mu^c = \nbl \cdot \bm{\epsilon}^c.
    \label{eq:charge-pd}
\)
The charge current does not flow across the surface
$j^c_z (\vr, t) \Big|_{z=0} = 0$, giving the Neumann boundary condition
$\sigma \partial_z \mu^c (\vr, t) \Big|_{z=0} 
= \epsilon^c_z (\vr,t) \Big|_{z=0}$.
Solving the differential equation yields
\(
   \mathbf{j}^c (\vr,t)
   = \bm{\epsilon}^c (\vr,t) 
   + \int_\Omega d^3 \vr' \, 
    \bm{\epsilon}^c  (\vr',t) \cdot  
    \nbl_{\vr'} \nbl_\vr \mathcal{G}_L(\vr,\vr') ,
    \label{eq:charge-current}
\)
where $\mathcal{G}_L(\vr,\vr')$ is the Green function for the Laplacian compatible with the Neumann boundary condition~\cite{suppmat}.\phantom{\cite{metal-review}} 

The magnetic field generated by the charge current is given by the Biot-Savart law,
\(
    \vB^c (\vr, t) 
    &= -\frac{e}{c}  \int_\Omega d^3 \vr' \, 
    \frac{\mathbf{j}^c (\vr',t) \times (\vr-\vr')} {|\vr-\vr'|^3},
    \label{eq:charge-field}
\)
where $c$ is the speed of light, and we have taken the charge carriers to have charge $-e$.
Inserting the magnetic field~(\ref{eq:charge-field}) into the definition~(\ref{eq:noise-correlation}), we obtain the magnetic noise due to charge fluctuations,
\(
    \tB^c_{ii'} (\omega)
    &=\frac{\pi e^2 k_B T \sigma}{c^2 d} \Lambda_{ii'},
    \label{eq:charge-noise}
\)
where $\Lambda = \text{diag}(1/2,1/2,1)$.  This reproduces exactly the result from the formulation in terms of transmission and reflection of electromagnetic fields at a metal surface~\cite{ford1984,henkel1999,henkel2005}, in the magnetostatic limit.

\sectionn{Valley.}|We next apply this approach to the valley degree of freedom. 
Different from the charge current, the valley fluctuations are neutral and thus compressible, which, as we have pointed out, turns out to be crucial for generating magnetic noise.
The weak intervalley scattering allows us to consider electron conservation in each valley separately: For the valley $p$ ($p = \pm$),
\(
    \partial_t \rho^p + \nbl \cdot \mathbf{j}^p = 0,
\)
where $\rho^p = (\nu/2) \mu^p$ and 
$\mathbf{j}^p = - (\sigma/2) \nbl \mu^p + \bm{\epsilon}^p$.
The total density of states $\nu$ at the Fermi surface includes both valleys, consistent with the convention of the total charge density
$\rho^c = \rho^+ + \rho^-$ and the average electrochemical potential
$\mu^c = (\mu^+ + \mu^-)/2$.
The Langevin noise here obeys
$\langle \epsilon^p_i (\vr, t) \epsilon^{p'}_{i'} (\vr', t') \rangle = \sigma k_B T \delta_{pp'} \delta_{ii'} \delta (\vr - \vr') \delta(t-t')$, supposing fluctuations in the two valleys are uncorrelated.
We therefore arrive at the stochastic diffusion equation
\(
    \partial_t \mu^p - D \nbl^2 \mu^p = -\frac{2}{\nu} \nbl \cdot \bm{\epsilon}^p,
    \label{eq:valley-pd}
\)
where $D = \sigma/\nu$ is the diffusion coefficient, as given by the Einstein relation. Under the boundary condition $j_z^p (\vr ,t) \Big|_{z=0} = 0$,
we solve for $\mu^p$ to obtain the currents
\(
    \mathbf{j}^p (\vr,t) 
    &= \bm{\epsilon}^p (\vr,t) - D
    \int_\Omega d^3 \vr' \int_{-\infty}^t  dt' \, \\
    &  \qquad \qquad \quad  \bm{\epsilon}^p  (\vr',t') \cdot  \nbl_{\vr'} \nbl_\vr  \mathcal{G}_D(\vr,\vr';t,t'),
    \label{eq:valley-current}
\)
where $\mathcal{G}_D(\vr,\vr';t,t')$ is the Green function for the diffusion equation satisfying the homogeneous boundary condition~\cite{suppmat}.

Our model dictates that the valley current~(\ref{eq:valley-current-spin}) generates magnetic fields equivalently to a local magnetization, via the demagnetization kernel,
\(
    \mathbf{B}^v(\vr, t) 
    =  \frac{g_v \mu_B}{v} \int_\Omega d^3 \vr' \left(-\nbl_{\vr} \nbl_{\vr'} \frac{1}{|\vr-\vr'|} \right) \cdot \mathbf{j}^v,
    \label{eq:valley-field}
\)
where $g_v$ is the effective g factor characterizing the coupling of the valley degree of freedom to an external magnetic field, and $\mu_B$ is the Bohr magneton. The valley contribution to the magnetic noise can then be calculated, yielding
\(
    \tB^v_{ii'} (\omega)
    &= \left(\frac{ g_v \mu_B}{v}\right)^2  
    \frac{4\pi k_B T \sigma}{d^3} \Lambda_{ii'} \int d\xi \,  \xi^2 e^{-2\xi} I^v(\xi,\zeta),
    \label{eq:valley-noise}
\)
with dimensionless quantities 
$\xi = Kd$, 
$\zeta = \omega d^2/D$, and 
$a = \sqrt{-i \omega/DK^2 +1} = \sqrt{-i \zeta/\xi^2 +1}$,
\(
    I^v(\xi,\zeta) = 
     1 + \frac{1+aa^*}{(a+a^*)aa^*} - \frac{1}{a^*} - \frac{1}{a}.
    \label{eq:valley-integrand}
\)
The integration is essentially taken over the magnitude of the wavevector $K = |\mathbf{K}|$ in the $xy$ plane, with $\xi^2 e^{-2\xi}$ as a form factor.
The anisotropy tensor $\Lambda$ is the same as in $\tB^c$~(\ref{eq:charge-noise}). 
The frequency dependence of the magnetic noise is contained in $I^v$, which scales as $\omega^2$ for $\omega \rightarrow 0$ and approaches $1$ for $\omega \rightarrow \infty$, $1-I^v \sim \omega^{-1/2}$.
The magnetic noise vanishes at zero frequency, consistent with our understanding that transverse components of the valley current do not contribute, in contrast to the charge current.

We estimate the magnitude of the valley contribution relative to the charge contribution,
\(
    \frac{\mathcal{B}^v_{zz}}{\mathcal{B}^c_{zz}}
    \sim g_v^2 
    \left(\frac{c \alpha}{v}\right)^2 
    \left(\frac{a_0}{d}\right)^2 \int d\xi \,  \xi^2 e^{-2\xi} I^v(\xi,\zeta),
\)
where we have reduced the result to physical constants, namely the fine-structure constant $\alpha$ and the Bohr radius $a_0$. 
The effective g factor $g_v$ can be greatly enhanced by the strong spin-orbit coupling, especially in topological semimetals~\cite{g-factor1960,g-factor1982,g-factor2015}. 
Taking  $g_v \sim 100$, the Fermi velocity $v \sim 10^5$~m/s~\cite{FermiVelocity}, the NV distance $d \sim 100$~nm, and evaluating the integral numerically at $\zeta = 5$,
we obtain $\mathcal{B}^v_{zz}/\mathcal{B}^c_{zz} \sim 0.2$. 
Recalling that $\mathcal{B}^c$~(\ref{eq:charge-noise}) 
is spectrally flat, in our treatment,
$\mathcal{B}^v$ can be easily recognized, for example, by taking a frequency derivative of the total measured noise $\partial_\omega \mathcal{B}$, exploiting the high frequency resolution~\cite{Yacoby2018} of NV probes. Moreover, this will not be spoiled by the presence of spin contribution, which 
will be seen to have a distinct
frequency behavior.

\begin{figure}[b]
    \centering
    \includegraphics[width = \linewidth]{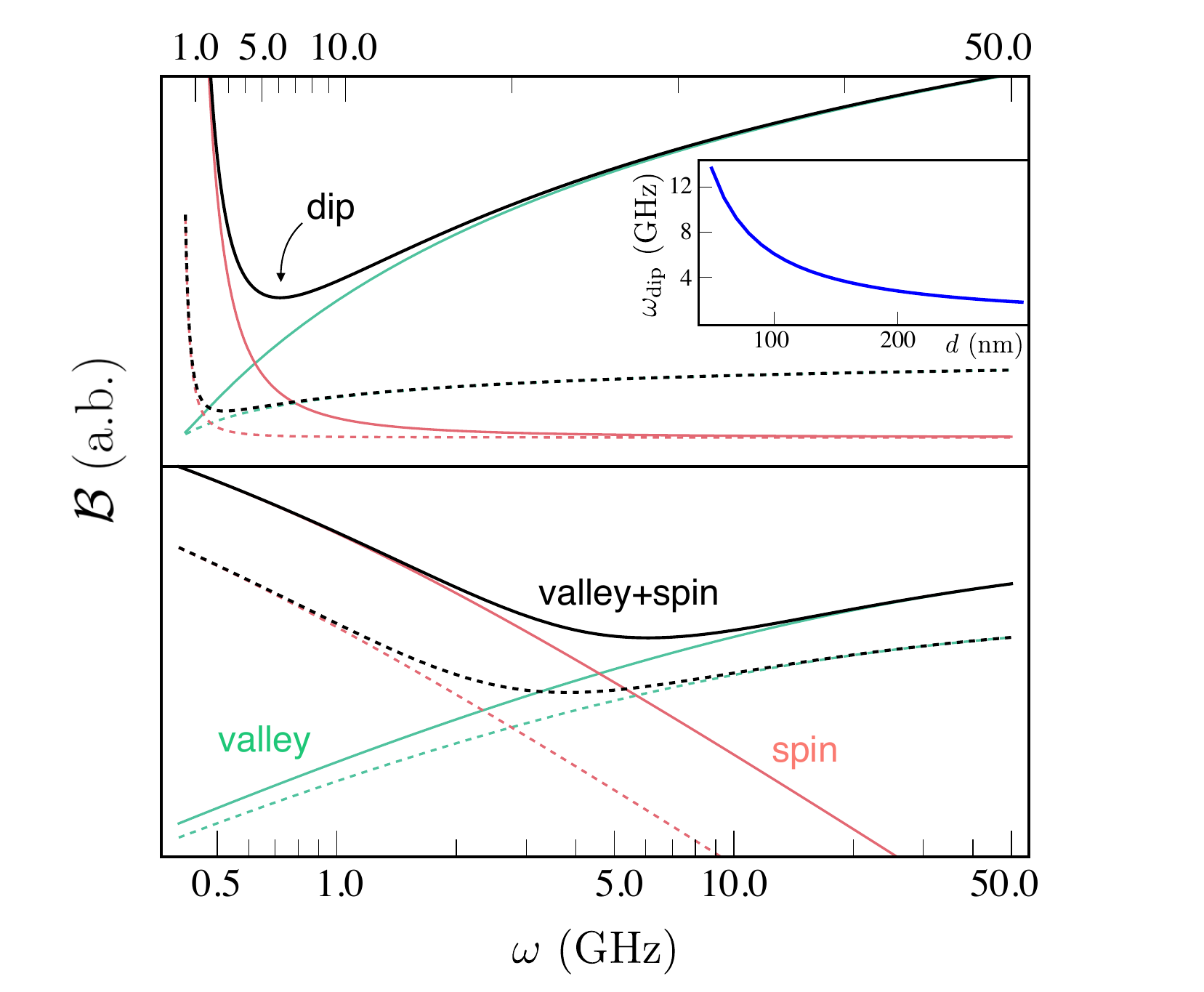}
    \caption{The frequency-dependent magnetic noise generated by valley (green) and spin (red) fluctuations in a linear (top) and a log-log (bottom) plot. The total magnetic noise (black) exhibits a ``dip'' feature at frequency $\omega_\text{dip}$. Two NV distances $d = 100$~nm (solid lines) and $d = 150$~nm (dashed lines) are plotted. Inset: $\omega_\text{dip}$ as a function of $d$. The vertical axis of the bottom plot spans three orders of magnitude (in arbitrary units).   }
    \label{fig:2}
\end{figure}

\sectionn{Spin.}|Fluctuations of the magnetic order, in the form of thermal magnon excitations, constitute a third ingredient of the magnetic noise in magnetic WSMs. 
Considering the magnon gap in magnetic WSMs can be several meV~\cite{Itoh2016,liu2021}, we focus on the subgap magnetic noise generated by the longitudinal (relative to the order parameter $\mathbf{M}$) spin fluctuations~\cite{Flebus2018,Hammel2020}.
The Langevin dynamics of the longitudinal spin density $s_z$ can be described by a stochastic diffusion equation, sometimes referred to as the Cahn-Hilliard model~\cite{CahnHilliard1958,ChaikinLubensky1995}:
\(
    \partial_t s_z - D^s \nbl^2 s_z = \epsilon^s,
    \label{eq:spin-pd}
\)
where $D^s$ is the spin diffusion constant. Here, the Gaussian white noise $\epsilon^s$ has the correlation function
$\langle \epsilon^s (\vr, t) \epsilon^s (\vr', t') \rangle 
= - 2 \sigma^s  k_B T  \nbl^2 \delta (\vr - \vr') \delta(t-t')$, where the spin conductivity $\sigma^s$ is related to the spin diffusion constant 
$D^s = \sigma^s/\chi_0$ 
by the static longitudinal spin susceptibility $\chi_0$. 
The corresponding boundary condition $j^s_z (\vr, t) \Big|_{z=0} = 0$ is specified in terms of the spin current 
$\mathbf{j}^s = -D^s \nbl s_z$. The solution is
\(
    s_z (\vr, t) = \int_\Omega d^3 \vr' \int_{-\infty}^t dt'
    \mathcal{G}_D^s (\vr, \vr';t,t') \, \epsilon^s (\vr',t'),
\)
where $\mathcal{G}_D^s (\vr, \vr';t,t')$ is simply the Green function of Eq.~(\ref{eq:valley-current}) with $D$ replaced by $D^s$.

The spin density generates demagnetization fields, and their the correlation function gives the magnetic noise
\(
    \mathcal{B}^s_{ii'} (\omega) 
    = \left(g_s \mu_B\right)^2 \frac{2 \pi  k_B T \chi_0}{\omega d^3} \Lambda_{ii'}
    \int d\xi \, \xi^2 e^{-2\xi} I^s(\xi, \eta),
    \label{eq:spin-noise}
\)
where $g_s$ is the g factor of localized spins, and
\(
    I^s (\xi, \eta) = -i  \left[ 
    \frac{1}{b(b+1)} - \frac{1}{b^*(b^*+1)}\right]
    \label{eq:spin-integral}
\)
where $b = \sqrt{-i \eta/\xi^2 +1}$, $\xi = Kd$ and $\eta = \omega d^2/D^s$. As $\omega \rightarrow 0$, $I^s$ scales as $\omega$ and $\mathcal{B}^s$ converges to a finite value. For $\omega \rightarrow \infty$, $I^s \sim 1/\omega$ and $\mathcal{B}^s$ vanishes asymptotically as $1/\omega^2$.

According to recent experiments around room temperature, the resultant NV transition rate from charge fluctuations in a good conductor~\cite{JohnsonNV2015} and that from longitudinal spin fluctuations in a magnetic insulator~\cite{AFMNV2020} are both of ms$^{-1}$ order. In magnetic WSMs with relatively low electrical conductivity, spin fluctuations may dominate the magnetic noise at low frequencies. However, the spin contribution quickly drops a few orders of magnitude with increasing frequency. Consequently, the total magnetic noise coming from valley and spin exhibits a ``dip'' feature in its frequency dependence, as shown in Fig.~\ref{fig:2}, on top of the flat charge contribution. The frequency $\omega_\text{dip}$ at the bottom of the dip shifts lower at a larger distance $d$.
For the plot, we have assumed similar charge and spin conductivities $\sigma \sim \sigma^s$~\footnote{The electric conductivity in a ferromagnetic WSM and the spin conductivity in a magnetic insulator can both be around $10^5 - 10^6$~S/m according to literature so far (see, for example,~\cite{cornelissen2015,AFMNV2020,hirschberger2016,Enke2018}).},
and taken both electron and magnon mean free paths $\sim 30$~nm, Fermi velocity $\sim10^5$~m/s and magnon velocity $\sim 10^3$~m/s, yielding $D \sim 10^{-3}$~m$^2$/s and $ D^s \sim 10^{-5}$~m$^2$/s.


\sectionn{Discussion.}|{It should be remarked that the valley noise derived here is based on a specific model~(\ref{eq:Hamiltonian}), as a demonstration of our formalism. The resulted prediction is not universal for all WSMs.} Different symmetry assumptions about spin and parity in the band model can lead to a different coupling scheme of the valley degree of freedom to magnetic fields. The frequency and distance dependence of the magnetic noise can therefore also reflect the applicability of the effective theories. 
{For specific materials, one may need to construct the coupling scheme from first-principles calculations.}

We have considered the magnetic noise primarily driven by the diffusive transport, neglecting the valley and spin relaxation effects. This applies to NV distances much smaller than the corresponding diffusion lengths (while larger than the mean free paths), which is often the case in a clean material. A finite valley relaxation time modifies the valley diffusion equation, and in our Langevin approach, is accompanied by an intervalley fluctuation of the axial charge density. Likewise, spin density relaxation and fluctuation can be added to the Cahn-Hilliard model. We refer to the discussion in the supplemental material~\cite{suppmat}. 
Furthermore, our treatment focuses on the low-frequency regime, where dynamic effects of the demagnetization fields, such as the Faraday induction, are negligible.

The Fermi arc states are universal in WSMs  and are generally dissipative~\cite{Kourtis-2016,Sukhachov2016}.
The magnetic noise they generate, which is not detected in our choice of geometry, may also be interesting to look into, by placing the NV near material surfaces parallel to the magnetic order $\mathbf{M}$.
The Fermi-arc contribution may be recognized by its distance dependence~\cite{Mintchev2019}, which is different from the bulk one.
For the spin channel, the change in the surface orientation relative to the order parameter would acquire a geometric factor~\cite{magnetostatic}. It would also be straightforward to account for an anisotropic electric conductivity under our framework for practical application.

{Going beyond the diffusive regime, a nanoscale probe such as NV can also probe the ballistic and perhaps the hydrodynamic transport of charge or spin, when the NV to material distance is smaller than or comparable to the electron or magnon mean free paths~\cite{JohnsonNV2015,fang2021magnon}. The spectral analysis of magnetic noise would also be particularly useful in the nonequilibrium regime, for example, in the presence of a steady spin or valley current, which may provide a cleaner detection of the corresponding nonlinear responses. } 

\begin{acknowledgements}

This work is supported by NSF under Grant No. DMR-2049979.
We thank Chunhui Rita Du for inspiring discussions.

\end{acknowledgements}

\bibliography{main}

\clearpage
\onecolumngrid

\setcounter{page}{1}

\begin{center}
  {\large \textbf{Supplemental Material for \\
  ``Flavors of Magnetic Noise in Quantum Materials''}} 
  \bigskip
  
  Shu Zhang and Yaroslav Tserkovnyak
  
  \small{\it{Department of Physics and Astronomy, University of California, Los Angeles, California 90095, USA}}
\end{center}

\appendix
\setcounter{equation}{0}
\renewcommand{\theequation}{S.\arabic{equation}}

\newcommand{\vjc}{{\mathbf{j}^c}}
\newcommand{\vjv}{{\mathbf{j}^v}}
\newcommand{\hz}{\hat{\mathbf{z}}}
\newcommand{\bk}{\bm{\kappa}}
\newcommand{\aak}{\textbf{\textkappa}}
\newcommand{\tb}{\widetilde{b}}
\newcommand{\ta}{\widetilde{a}}
\renewcommand{\v}[1]{\mathbf{#1}}

In this document, we present the details on the generation of the magnetic noise by the fluctuations of the charge, valley, and spin degrees of freedom in a magnetic Weyl semimetal, via different generating kernels.
Specifically, we look into the magnetic field $\vB (\vr_\text{NV}, t)$ at the position of the NV center $\vr_\text{NV} = (0,0,d)$ generated by the material occupying the half space $\Omega$ with $z \le 0$, assuming the material dimensions are much larger than $d$.

The following identity and its variants will be frequently used, which is convenient for the broken translational invariance in the $z$ direction,
\(
    \frac{1}{|\vr-\vr'|} = \int \frac{d^2 \vK}{(2\pi)^2}  \, \frac{2 \pi}{K} e^{i \vK \cdot (\vR-\vR')-K|z-z'|}, 
\)
where $\vR = (x,y,0)$, and $\vK = (k_x, k_y, 0)$ denotes the wavevector in the $x$ and $y$ directions, $K =|\vK|$.

\section{I. Transport}

\subsection{A. Charge}

We take a stochastic approach and study the response of the charge current to a Langevin noise, and verify that our method reproduces the known result for the Johnson-Nyquist noise from the thermal fluctuations of electric charge current in a metal.
The charge transport obeys the continuity equation
\(
    \partial_t \rho^c + \nbl \cdot \vjc = 0.
\)
In the low frequency limit, the charge density fluctuation is efficiently screened by Coulomb interactions. We thus consider the transverse current fluctuations only, $\nbl \cdot \vjc = 0$, where $\vjc = - \sigma \nbl \mu^c + \bm{\epsilon}^c$. Here $\sigma$ is the conductivity, $\mu^c$ is the chemical potential,
and $\bm{\epsilon}^c$ is a Gaussian white noise satisfying 
$\langle \epsilon^c_i (\vr, t) \rangle = 0$ and 
$\langle \epsilon^c_i (\vr, t) \epsilon^c_{i'} (\vr', t') \rangle =  2 \sigma k_B T \delta_{ii'} \delta (\vr - \vr') \delta(t-t')$, as a result of the fluctuation dissipation theorem at high temperatures. 
We find the response of the charge current $\vjc$ to the thermal noise $\bm{\epsilon}^c$ by solving the following differential equation
\(
    \sigma \nbl^2 \mu^c = \nbl \cdot \bm{\epsilon}^c.
    \label{eq:charge-pd-suppmat}
\)
In the free space, it is easy to show by the Fourier transform that
$-\sigma \vk^2 \mu^c = i \vk \cdot \bm{\epsilon}^c$ and 
$\vjc = \bm{\epsilon}^c - \vk \vk \cdot \bm{\epsilon}^c / k^2$,
which gives the familiar transverse current correlation function
\(
\langle j^c_i (\vr, t) j^c_{i'} (\vr', t') \rangle 
= \left[\delta_{ii'} \delta (\Delta \mathbf{r}) + \frac{\delta_{ii'} |\Delta \mathbf{r}|^2 - 3 \Delta \mathbf{r}_i \Delta \mathbf{r}_{i'}}{4\pi |\Delta \mathbf{r}|^5} \right] 2 \sigma k_B T   \delta(t-t'),
\)
where $\Delta \mathbf{r} = \mathbf{r} - \mathbf{r}'$.

Here, since the current does not flow out of the material, we solve Eq.~(\ref{eq:charge-pd-suppmat}) in the half space $\Omega$ with the Neumann boundary condition 
$\sigma \partial_z \mu^c (\vr, t) \Big|_{z=0} 
= \epsilon^c_z (\vr,t) \Big|_{z=0}$.
The Green function for the Laplacian with the homogeneous Neumann boundary condition is [the time dependent factor is simply $\delta(t-t')$] 
\(
    \mathcal{G}_L(\vr,\vr') 
    = -\frac{1}{4\pi}\left[ \frac{1}{\sqrt{(x-x')^2 + (y-y')^2 + (z-z')^2}} +\frac{1}{\sqrt{(x-x')^2 + (y-y')^2 + (z+z')^2}}
    \right] .
    \label{eq:charge-Green}
\)
Therefore,
\(
    \mu^c(\vr,t) 
    &= \int_\Omega d^3 \vr'  \, \mathcal{G}_L(\vr,\vr') 
    \frac{\nbl_{\vr'} \cdot \bm{\epsilon}^c  (\vr',t)}{\sigma}
    - \int_{\partial \Omega} d^2 \vr'\, \mathcal{G}_L(\vr,\vr') \frac{\epsilon^c_z (\vr',t)}{\sigma}  ,\\
\)
and integrating by parts
    \(
    \vjc (\vr,t) 
    &= \bm{\epsilon}^c (\vr,t) + 
    \int_\Omega d^3 \vr' \, 
    \bm{\epsilon}^c  (\vr',t) \cdot  \nbl_{\vr'} \nbl_\vr \mathcal{G}_L(\vr,\vr').  \\
    \)
Using the Fourier transform of the Green function (\ref{eq:charge-Green}) in the $x$ and $y$ directions with $\vR = (x,y,0)$,
\(
     \mathcal{G}_L(\vK,z,z') = 
     -\frac{1}{2K}
    \left( e^{-K |z-z'|}+e^{-K|z+z'|} \right),
\)
we take the Fourier transfrom of $\vjc(\vr,t)$ in the following form,
\(
    \vjc (\vK,t) 
    & = \int_{-\infty}^0 dz \, e^{Kz} \int d^2 \vR \, e^{-i \vK \cdot \vR} \vjc (\vr,t)\\
    &= \int_{-\infty}^0 d z \, e^{Kz} 
    \int d^2  \vR \, e^{-i \vK \cdot \vR} \, 
    \left[ \bm{\epsilon}^c(\vr,t) -   \epsilon_z^c(\vr,t) \hz \right]\\
    & \qquad \qquad - \frac{1}{2K}
    \int_{-\infty}^0 d z' \, e^{Kz'}  
    \int d^2 \vR'  \, e^{-i \vK \cdot \vR'} \bm{\epsilon}^c  (\vr',t) \cdot
     \left( \vK \bm{\kappa}^*
    \frac{1}{K} - \bm{\kappa} \bm{\kappa} z'  \right),
\)
where $\bm{\kappa} = \vK + i K \hz$.
The second term in the square brackets comes from the Dirac $\delta$-function in $\nbl_{\vr'} \nbl_\vr G^c(\vr,\vr')$. 

In the magnetostatic limit, where the wavelength of the electromagnetic field and the skin depth are both much larger than the NV distance $d$, the Biot-Savart law gives the magnetic field generated by the electric charge current distribution,
\(
    \vB^c (\vr_\text{NV}, t) 
    &= -\frac{e}{c}  \int_\Omega d^3 \vr \, 
    \vjc (\vr,t) \times  \nbl_\vr \frac{1}{|\vr_\text{NV}-\vr|} \\
    &= -\frac{e}{c} \int \frac{d^2 \vK}{(2\pi)^2} \,  
    \frac{2\pi}{K}  e^{-Kd}
    (i \bk) \times  \vjc (\vK,t)\\
    &= -\frac{ e}{c} \int \frac{d^2 \vK}{(2\pi)^2} \,  
    \frac{2\pi}{K}  e^{-Kd}
    \left[ i \bk \times \bm{\epsilon}^c(\vK,t) + \frac{1}{K} (\hz \times \vK)  \bm{\epsilon}^c(\vK,t) \cdot \bk \right],\\
    \label{eq:charge-field-supmatt}
    \)
where $\bm{\epsilon}^c (\vK,t) = \int_{-\infty}^0 dz \, e^{Kz} \int d^2 \vR \, e^{-i \vK \cdot \vR} \, \bm{\epsilon}^c (\vr,t)$. We have used 
$\bm{\kappa} \times \bm{\kappa}^*  = i2K \v{z} \times \vK$ and 
$\bm{\kappa} \times \bm{\kappa}  = 0$. This result reproduces that in Ref.~\cite{henkel2005} derived from Maxwell equations with proper boundary conditions, and is consistent with the approach of finding the Fresnel coefficients~\cite{ford1984,henkel1999}.
Finally, we obtain the magnetic noise tensor contributed from charge fluctuations, according to Eq.~(\ref{eq:noise-correlation}),
\(
    \tB^c_{ii'} (\omega)
    &= \left(\frac{ e}{c}\right)^2 \int d^2 \vK  \frac{2\sigma k_B T}{2K^3} e^{-2K d}
    \left(
    \epsilon^{ijk} \epsilon^{i' j' k } \kappa_j \kappa_{j'}^*
    + \frac{i}{K} \epsilon^{ijk} \epsilon^{i'zk'} \kappa_j K_{k'} \kappa_{k}^*
    - \frac{i}{K} \epsilon^{i'j'k'} \epsilon^{izk} \kappa^*_{j'} K_{k} \kappa_{k'}
    + 2\epsilon^{izk} \epsilon^{i'zk'}K_{k}K_{k'}
    \right)\\
    &= \left(\frac{ e}{c}\right)^2 \int d K
     e^{-2K d} 2 \pi \sigma k_B T  \Delta_{ii'} \\
    &=\frac{\pi e^2 k_B T \sigma}{c^2 d} \Delta_{ii'},
\)
where the tensor $\Delta = \text{diag}(1/2,1/2,1)$ and Einstein summation is taken over repeated indices. 

\subsection{B. Valley}

Following the same approach, we next look into the contribution from the valley current. We first focus on the particle conservation in a single valley $p$, assuming the interaction and scattering between the two Weyl nodes are weak. We do not consider the chiral-anomaly related perturbations here. 
\(
    \partial_t \rho^p + \nbl \cdot \v{j}^p = 0,
\)
where 
$\rho^p = (\nu/2) \mu^p$ and 
$\v{j}^p = - (\sigma/2) \nbl \mu^p + \bm{\epsilon}^p$. Here $\nu$ refers to the total density of states at the Fermi surface, including both valleys. 
Assuming no correlation between $\bm{\epsilon}^+$ and $\bm{\epsilon}^-$, $\langle \epsilon^p_i (\vr, t) \epsilon^{p'}_{i'} (\vr', t') \rangle = \sigma k_B T \delta_{pp'} \delta_{ii'} \delta (\vr - \vr') \delta(t-t')$.
The differential equation to be solved is then
\(
    \partial_t \mu^p - D \nbl^2 \mu^p = -\frac{2}{\nu} \nbl \cdot \bm{\epsilon}^p
\)
where the diffusion coefficient $D = \sigma/\nu$, as is consistent with the Einstein relation and the Drude law.  The boundary condition $j_z^p (\vr ,t) \Big|_{z=0} = 0$ is similar to the case with charge. We use the Green function for the diffusion equation satisfying the homogeneous Neumann boundary condition, and its Fourier transform
\(
    &\mathcal{G}_D (\vr, \vr'; t,t') 
    = \frac{\Theta (t-t')}{[4 \pi D (t-t')]^{3/2}} e^{-[(x-x')^2 + (y-y')^2]/4D(t-t')}
    \left( e^{-(z-z')^2/4D(t-t')} + e^{- (z+z')^2/4D(t-t')} \right),\\
    &\mathcal{G}_D (\vK, z, z'; \omega) =   \frac{1}{2D a K} 
    \left( e^{-a K|z-z'|}  + e^{-aK|z+z'|} \right),
    \label{eq:valley-Green}
\)
where $\Theta (t-t')$ is the Heaviside step function. Here, the dimensionless parameter $a = \sqrt{-i\omega/DK^2 + 1}$ carries the frequency dependence in the following context, and is evaluated in the branch with $\text{Re} \,a > 0$. 
The current distribution is therefore
\( \label{eq:valley-current-suppmat}
    \v{j}^p (\vr,t) 
    &= \bm{\epsilon}^p (\vr,t) - D
    \int_\Omega d^3 \vr' \int_{-\infty}^t  dt' \, \bm{\epsilon}^p  (\vr',t') \cdot \nbl_{\vr'}  \nbl_\vr \mathcal{G}_D(\vr,\vr';t,t'), \\
\)
and
\(
    \v{j}^p (\vK, \omega) 
    & = \int_{-\infty}^0 dz \, e^{Kz} \int d^2 \vR \int dt \,
    e^{i\omega t -i \vK \cdot \vR} \v{j}^p (\vr,t)\\
    &= \int_{-\infty}^0 dz \, e^{Kz} \int d^2 \vR \int dt \,
    e^{i\omega t - i \vK \cdot \vR} \, 
    \left[ \bm{\epsilon}^p (\vr,t) - {\epsilon}_z^p (\vr,t) \hz\right] \\
    &- \frac{1}{2 a K^2}  \int_{-\infty}^0 dz' \int d^2 \vR'  
    \int dt \,  e^{i \omega t' - i \vK \cdot \vR'}\, 
    \bm{\epsilon}^p  (\vr',t') \cdot 
    \left[ {\aak}^\star {\aak}^\star \frac{e^{Kz'}}{a+1} 
    + \aak \aak \frac{e^{Kz'} - e^{aK z'}}{a-1} 
    +\aak {\aak}^\star \frac{e^{a K z'}}{a+1} \right],\\
\)
where 
$\aak = \vK + i a K  \hz$, and $\aak^\star = \vK - i a K  \hz$. 

The valley current density behaves like a local magnetization in generating the magnetic field:
\(
    \vB^v (\vr_\text{NV}, \omega) 
    &=  \frac{g_v \mu_B}{v} \int_\Omega d^3 \vr \left(-\nbl_{\vr_\text{NV}} \nbl_{\vr} \frac{1}{|\vr_\text{NV}-\vr|} \right) \cdot \vjv (\vr, \omega)\\
    &= -\frac{g_v \mu_B}{v}  \int \frac{d^2 \vK}{(2\pi)^2} \,  
    \frac{2\pi}{K}  e^{-Kd}
    \bk \bk \cdot  \left[ \v{j}^+ (\vK,\omega) - \v{j}^- (\vK,\omega) \right] \\
    &=  -\frac{g_v \mu_B}{v}  \int \frac{d^2 \vK}{(2\pi)^2} \,  
    \frac{2\pi}{K}  e^{-Kd}
    \int_{-\infty}^0 dz \int d^2 \vR  \int dt \,
    e^{i\omega t - i \vK \cdot \vR} \, 
    \bk \left( e^{Kz} \bk - \frac{e^{aKz}}{a} \aak \right) \cdot
    \left[ \bm{\epsilon}^+ (\vr,t) - \bm{\epsilon}^- (\vr,t) \right],
\)
where $g_v$ is the effective g factor describing the coupling of the valley degrees of freedom to an external magnetic field and $v$ is the Fermi velocity. We have used $\bk \cdot \aak = - K^2 (a-1)$ and $\bk \cdot \aak^\star = K^2 (a + 1)$.
With the help of 
\(
    &\Big\langle 
    \int_{-\infty}^0  dz \, e^{c_1 z} \int d^2  \vR \int dt \, e^{-i\omega t + i \vK \cdot \vR} \, 
    \epsilon^p_i(\vr ,t)  
    \int_{-\infty}^0  dz' \, e^{{c_2}^* z'}\int d^2 \vR'  \int dt' \, e^{i \omega' t' -i \vK' \cdot \vR'} 
    \epsilon^{p'}_{i'}  (\vr', t') \Big\rangle\\
    & = \frac{\sigma k_B T}{c_1 + {c_2}^*}  
    (2\pi)^3 \delta_{pp'} \delta_{ii'} \delta (\vK - \vK') \delta(\omega - \omega'),
\)
we obtain the magnetic noise tensor generated by valley fluctuations
\(\label{eq:valley-noise-supp}
    \tB^v_{ii'} (\omega)
    &= \frac{1}{2\pi} \langle B_i (\vr_\text{NV},\omega) B_{i'}^* (\vr_\text{NV},\omega) \rangle \\
    &= \left(\frac{ g_v \mu_B}{v}\right)^2 \int \frac{d^2 \vK}{(2\pi)^2} 
    \frac{(2\pi)^2}{K} e^{-2K d} 2 \sigma k_B T \kappa_i  {\kappa}_{i'}^* 
    \left[ 1 + \frac{1+aa^*}{(a+a^*)aa^*} - \frac{1}{a^*} - \frac{1}{a}\right]\\
    &= \left(\frac{ g_v \mu_B}{v}\right)^2  
    \frac{4\pi k_B T \sigma}{d^3} \Lambda_{ii'} \int d\xi \,  \xi^2 e^{-2\xi} I^v (\xi, \zeta),
\)
where
\(
    I^v (\xi, \zeta) = 1 + \frac{1+aa^*}{(a+a^*)aa^*} - \frac{1}{a^*} - \frac{1}{a},
\)
with dimensionless quantities 
$\xi = Kd$, 
$\zeta = \omega d^2/D$, and $a = \sqrt{-i \zeta/\xi^2 +1}$.
The anisotropy tensor $\Delta$ is the same as in $\tB^c$.

\subsection{C. Spin}

We invoke the Cahn-Hilliard model~\cite{CahnHilliard1958} to describe the  Langevin dynamics of the longitudinal spin density $s_z$
\(
    \partial_t s_z - D^s \nbl^2 s_z = \epsilon^s,
\)
where $D^s$ is the spin diffusion constant and $\epsilon^s$ is a scalar Gaussian white noise with the correlation function
$\langle \epsilon^s (\vr, t) \epsilon^s (\vr', t') \rangle 
= - 2 \sigma^s  k_B T  \nbl^2 \delta (\vr - \vr') \delta(t-t')$, containing the Laplacian operator. The spin conductivity $\sigma^s$ is related to the spin diffusion constant 
$D^s = \sigma^s/\chi_0$ 
by the static longitudinal spin susceptibility $\chi_0$. 
The boundary condition $j^s_z (\vr, t) \Big|_{z=0} = 0$ is specified in terms of the spin current 
$\mathbf{j}^s = -D^s \nbl s_z$. The solution is
\(
    s_z (\vr, t) = \int_\Omega d^3 \vr' \int_{-\infty}^t 
    \mathcal{G}_D^s (\vr, \vr';t,t') \, \epsilon^s (\vr',t').
\)
The Green function $\mathcal{G}_D^s (\vr, \vr';t,t')$ is simply  Eq.~(\ref{eq:valley-Green}) with $D \rightarrow D^s$,
\(
    &\mathcal{G}_D^s (\vr, \vr'; t,t') 
    = \frac{\Theta (t-t')}{[4 \pi D^s (t-t')]^{3/2}} e^{-[(x-x')^2 + (y-y')^2]/4D^s(t-t')}
    \left( e^{-(z-z')^2/4D^s(t-t')} + e^{- (z+z')^2/4D^s(t-t')} \right)\\
    &\mathcal{G}_D^s (\vK, z, z'; \omega)  = \frac{1}{2D b K} 
    \left( e^{-b K|z-z'|}  + e^{-b K|z+z'|} \right),
    \label{eq:spin-Green}
\)
where $b = \sqrt{-i\omega/D^s K^2 +1}$.
Taking the Fourier transform,
\(
    s_z (\vK, \omega) 
    & = \int_{-\infty}^0 dz \, e^{Kz} \int d^2 \vR \int dt \,
    e^{i\omega t -i \vK \cdot \vR} s_z (\vr,t)\\
    &= \int_\Omega d^3 \vr' \int_{-\infty}^t dt'
     e^{i \omega t' - i \vK \cdot \vR'}\,
    \frac{1}{D^s b K^2}  
    \frac{b e^{K z'} - e^{bKz'}}{(b+1)(b-1)} \epsilon^s (\vr',t') 
\)

The spin density generates stray fields via the demagnetization kernel,
\(\label{eq:B-sz-suppmat}
    \v{B}^s(\vr_\text{NV},\omega) 
    &= g_s \mu_B \int_\Omega d^3 \vr \left(-\nbl_{\vr_\text{NV}} \nbl_{\vr} \frac{1}{|\vr_\text{NV}-\vr|} \right) \cdot  s_z(\vr,\omega) \hz\\
    &= g_s \mu_B  \int \frac{d^2 \vK}{(2\pi)^2} \,  
    \frac{2\pi}{K}  e^{-Kd}
    \bk \int_{-\infty}^0 dz\int d^2 \vR \int_{-\infty}^t dt
     e^{i \omega t - i \vK \cdot \vR}\,
    \frac{1}{iD^s b K}  
    \frac{b e^{K z} - e^{bKz}}{(b+1)(b-1)} \epsilon^s (\vr,t).
\)
Using the correlation relation of the Gaussian noise, we obtain the magnetic noise tensor
\(
    \mathcal{B}^s_{ii'} (\omega) 
    &= (g_s \mu_B)^2 \int \frac{d^2 \vK}{(2\pi)^2} \,  
    \frac{(2\pi)^2}{K}  e^{-2Kd} \kappa_i \kappa^*_{i'} 
    \frac{\sigma^s k_B T}{ i D^s \omega}
    \left[ \frac{1}{b(b+1)} - \frac{1}{b^*(b^*+1)} \right]\\
    &= \left(g_s \mu_B\right)^2 \frac{2 \pi  k_B T \chi_0}{\omega d^3} \Lambda_{ii'}
    \int d\xi \, \xi^2 e^{-2\xi} I^s(\xi, \eta),
     \label{eq:spin-noise-suppmat}
\)
where $g_s$ is the g factor of localized spins, and
\(
     I^s(\xi, \eta) = -i 
    \left[ \frac{1}{b(b+1)} - \frac{1}{b^*(b^*+1)}\right]
\)
with $\xi = Kd$, $\eta = \omega d^2/D^s$ and $b = \sqrt{-i \eta/\xi^2 +1}$. 

\section{II. Relaxation}

\subsection{A. Spin}

In the Langevin approach, the spin relaxation can be handled by replacing the correlation functions of the Gaussian noise by 
$\langle \widetilde{\epsilon}^s (\vr, t) \widetilde{\epsilon}^s (\vr', t') \rangle 
= 2  k_B T \chi_0  (1/\tau^s -D^s \nbl^2 ) \delta (\vr - \vr') \delta(t-t')$ in
\(
    \left(\partial_t + \frac{1}{\tau^s} \right) s_z - D^s \nbl^2 s_z 
    = \widetilde{\epsilon}^s,
\)
where $\tau^s$ is the spin relaxation time.
This is because the consideration of the spin relaxation has to be accompanied by a spin density fluctuation, as dictated by the fluctuation-dissipation theorem. 
Here, the Green function and its Fourier transform are, respectively,
\(
    &\widetilde{\mathcal{G}}^s_D (\vr, \vr'; t,t') 
    = \frac{\Theta (t-t')}{[4 \pi D^s (t-t')]^{3/2}} e^{-[(x-x')^2 + (y-y')^2]/4D^s(t-t')}
    \left( e^{-(z-z')^2/4D^s(t-t')} + e^{- (z+z')^2/4D^s(t-t')} \right)
    e^{-(t-t')/\tau^s},\\
    & \widetilde{\mathcal{G}}^s_D (\vK, z, z'; \omega) 
    = \frac{1}{2D^s \tb  K } 
    \left( e^{-\tb K|z-z'|}  + e^{-\tb K|z+z'|} \right),
\)
where $\tb  = \sqrt{(-i\omega + 1/\tau^s)/D^s K^2 +1}$. 
The magnetic noise thus becomes
\(  \label{eq:spin-noise-supp}
    \widetilde{\mathcal{B}}^s_{ii'} (\omega) 
    &= (g_s \mu_B)^2 \int \frac{d^2 \vK}{(2\pi)^2} \,  
    \frac{(2\pi)^2}{K^3}  e^{-2Kd} \kappa_i \kappa^*_{i'} 
    \frac{ 2 k_B T \chi_0}{ D^s}  \frac{ (1+\tb+\tb^*)(\tb+1)(\tb^*+1)+[2(\tb+\tb^*+1)^2 + \tb \tb^* (\tb+\tb^*)]/2K^2D^s \tau^s}{\tb \tb^* (\tb+1)^2(\tb^*+1)^2(\tb+\tb^*)}\\
    &= (g_s \mu_B)^2  \frac{\pi k_B T \chi_0}{  D^s}
   \Delta_{ii'} \int dK e^{-2K d}\,
     \left[\frac{2+\tb }{\tb (\tb +1)^2} + \frac{2+\tb^* }{\tb^* (\tb^* +1)^2}\right].
\)
In the limit $1/\tau^s \rightarrow 0$, $\text{Re} \left[ (2+\tb )/\tb (\tb +1)^2 \right] \rightarrow \text{Im}\left[(2D^sK^2/\omega)/b(b+1)\right]$, we arrive at the same result as in Eq.~(\ref{eq:spin-noise-suppmat}). In the opposite limit $\tau^s \rightarrow 0$, $(2+\tb )/\tb (\tb +1)^2 \approx D^s K^2 \tau^s$, $\tB^s_{ii'} (\omega) \rightarrow \left(g_s \mu_B\right)^2 \pi k_B T\chi_0 \tau^s \Delta_{ii'}/2d^3$ is independent of $D^s$, as expected from the relaxation dominated equation of motion.

Alternatively, the noise can be computed from the dynamic spin susceptibility, with the help of the fluctuation dissipation theorem. We derive dynamic spin susceptibility from the diffusion equation for the longitudinal spin dynamics
\(
    \partial_t s_z + \nbl \cdot \mathbf{j}^s = 
    - \frac{\chi_0}{\tau^s} \mu^s,
    \label{eq:spin-diffusion}
\)
where 
$\mathbf{j}^s = -\sigma^s \nbl \mu^s $ is the spin current, $\mu^s = s_z/\chi_0 -  h$ is the spin chemical potential, and $h$ is a force thermodynamically conjugate to $s_z$, given by the external magnetic field. The spin distributionr relaxes towards $\chi_0 h$ ($\mu^s = 0$ in equilibrium), characterized by the spin relaxation time $\tau^s$.
The differential equation
\(
    \left( \partial_t + \frac{1}{\tau^s} \right)
    \mu^s - D^s \nbl^2 \mu^s 
    = -\partial_t  h.
\)
again has the boundary condition $\partial_z \mu^s (\vr,t) \Big|_{z=0} = 0$. 
From the solution of $\mu^s (\vr,t)$, we have
\(
    s_z (\vr,t) 
    &= \chi_0 \int_\Omega d^3 \vr' \int_{-\infty}^t dt' \, 
    \left[\delta(\vr - \vr') \delta(t-t') - \widetilde{\mathcal{G}}_D^s (\vr, \vr'; t,t')
    \partial_{t'}   \right] h (\vr', t') ,
\)
which defines the longitudinal susceptibility $\chi (\vr, \vr'; t,t')$. 
The Fourier transform of the dynamic spin susceptibility is therefore
\(
    \chi(z,z',\vK,\omega) 
    = \chi_0 \left[\delta(z-z')  + \frac{i \omega}{2D^s \tb  K}
    \left( e^{-\tb K|z-z'|}  + e^{-\tb K|z+z'|} \right)\right].
\)
The spin induced magnetic noise is then
\(
    \widetilde{\tB}^s_{ii'} (\omega)
    &= \left(g \mu_B\right)^2 \int \frac{d^2 \vK}{(2\pi)^2} 
    \frac{(2\pi)^2}{K^2} e^{-2K d}
    \kappa_i  {\kappa}_{i'}^* \kappa_j  {\kappa}_{j'}^*
    \int_{-\infty}^0 dz \int_{-\infty}^0 dz' \, e^{K(z+z')}\,
    \frac{2k_B T}{\omega}   \text{Im} \chi(z,z',\vK,\omega)\\
    &= \left(g \mu_B\right)^2 \frac{2\pi k_B T\chi_0}{D^s}  
    \Delta_{ii'}
    \int dK e^{-2K d}\, \text{Re} \left[ \frac{2+\tb }{\tb (\tb +1)^2}\right],
\)
which reproduces the result~(\ref{eq:spin-noise-suppmat}) from the Langevin approach.

\subsection{B. Valley}

In the same spirit, we consider the following diffusion and relaxation of the valley degree of freedom 
\(
    \left( \partial_t +\frac{1}{\tau^v} \right) \rho^v + \nbl \cdot \v{j}^v = \widetilde{\epsilon}^v,
\)
where
$\rho^v = (\nu/2) \mu^v$, $\nu$ is the total density of states at the Fermi surface, $\mu^v = (\mu^+ - \mu^-$)/2 is the valley chemical potential,
$\v{j}^v = - (\sigma/2) \nbl \mu^v + \bm{\epsilon}^v$, and $\tau^v$ is the valley relaxation time. The intravalley noise has the same correlation as before $\langle \epsilon^v_i (\vr, t) \epsilon^{v}_{i'} (\vr', t') \rangle = 2\sigma k_B T \delta_{ii'} \delta (\vr - \vr') \delta(t-t')$, and we also account for an intervalley fluctuation of the axial charge density $\langle \widetilde{\epsilon}^v (\vr,t) \rangle =0$ and $\langle \widetilde{\epsilon}^v (\vr,t) \widetilde{\epsilon}^v (\vr',t') \rangle = (\nu/\tau^v)2 k_B T \delta (\vr - \vr') \delta(t-t')$.
The differential equation to be solved is then
\(
    \left( \partial_t +\frac{1}{\tau^v} \right)  \mu^v - D \nbl^2 \mu^v 
    = \frac{2}{\nu} ( - \nbl \cdot \bm{\epsilon}^v + \widetilde{\epsilon}^v),
\)
with the Neumann boundary condition. The Green function in this case is
\(
    &\widetilde{\mathcal{G}}_D (\vr, \vr'; t,t') 
    = \frac{\Theta (t-t')}{[4 \pi D (t-t')]^{3/2}} e^{-[(x-x')^2 + (y-y')^2]/4D(t-t')}
    \left( e^{-(z-z')^2/4D(t-t')} + e^{- (z+z')^2/4D(t-t')} \right)
    e^{-(t-t')/\tau^s},\\
    & \widetilde{\mathcal{G}}_D (\vK, z, z'; \omega) 
    = \frac{1}{2D \ta  K } 
    \left( e^{-\ta K|z-z'|}  + e^{-\ta K|z+z'|} \right),
\)
where $\ta = \sqrt{(-i\omega + 1/\tau^v)/DK^2 + 1}$. Following Eqs.~(\ref{eq:valley-current-suppmat}-\ref{eq:valley-noise-supp}),
\(
    \widetilde{\v{j}}^v (\vr,t) 
    &= \bm{\epsilon}^v (\vr,t) - D
    \int_\Omega d^3 \vr' \int_{-\infty}^t  dt' \, \left[ \, \widetilde{\epsilon}^v  (\vr',t') + \bm{\epsilon}^v  (\vr',t') \cdot \nbl_{\vr'} \right] \nbl_\vr \widetilde{\mathcal{G}}_D(\vr,\vr';t,t'), \\
\)
\(
    \widetilde{\v{j}}^v (\vK, \omega) 
    &= \int_{-\infty}^0 dz \, e^{Kz} \int d^2 \vR \int dt \,
    e^{i\omega t - i \vK \cdot \vR} \, 
    \left[ \bm{\epsilon}^v (\vr,t) - {\epsilon}_z^v (\vr,t) \hz\right] \\
    &- \frac{1}{2 \ta K^2}  \int_{-\infty}^0 dz' \int d^2 \vR'  
    \int dt \,  e^{i \omega t' - i \vK \cdot \vR'}\, 
    \bm{\epsilon}^v  (\vr',t') \cdot 
    \left( {\widetilde{\aak}}^\star {\widetilde{\aak}}^\star \frac{e^{Kz'}}{\ta+1} 
    + \widetilde{\aak} \widetilde{\aak} \frac{e^{Kz'} - e^{\ta K z'}}{\ta-1} 
    +  \widetilde{\aak} {\widetilde{\aak}}^\star \frac{e^{\ta K z'}}{\ta+1} \right)\\
    &- \frac{i}{ 2 \ta K^2}  \int_{-\infty}^0 dz' \int d^2 \vR'  
    \int dt \,  e^{i \omega t' - i \vK \cdot \vR'}\, 
    \widetilde{\epsilon}^v  (\vr',t') 
    \left( \widetilde{\aak} \frac{e^{Kz'} - e^{\ta Kz'}}{\ta-1} 
    + \widetilde{\aak}^\star \frac{e^{Kz'} + e^{\ta Kz'}}{\ta+1} \right),
\)
where $\widetilde{\aak} = \vK + i \ta K \hz$ and $\widetilde{\aak}^\star = \vK - i \ta K \hz$.
\(
    \widetilde{\vB}^v (\vr_\text{NV}, \omega) 
    &= -\frac{g_v \mu_B}{v}  \int \frac{d^2 \vK}{(2\pi)^2} \,  
    \frac{2\pi}{K}  e^{-Kd}
    \bk \bk \cdot  \widetilde{\v{j}}^v  (\vK,\omega)  \\
    &=  \frac{g_v \mu_B}{v}  \int \frac{d^2 \vK}{(2\pi)^2} \,  
    \frac{2\pi}{K}  e^{-Kd}
    \int_{-\infty}^0 dz \int d^2 \vR  \int dt \,
    e^{i\omega t - i \vK \cdot \vR} \, 
    \bk
    \left[
    \left(  - e^{Kz} \bk  + \frac{e^{ \ta Kz}}{\ta} \widetilde{\aak} \right ) \cdot \bm{\epsilon}^v (\vr,t) 
    +\frac{i e^{\ta K z}}{\ta}  \widetilde{\epsilon}^v  (\vr',t')\right].
\)
The magnetic noise turns out to be
\(
    \widetilde{\tB}^v_{ii'} (\omega) 
    &= \left(\frac{ g_v \mu_B}{v}\right)^2 \int \frac{d^2 \vK}{(2\pi)^2} 
    \frac{(2\pi)^2}{K} e^{-2K d} (2 \sigma k_B T) \kappa_i  {\kappa}_{i'}^* 
    \left[  \frac{(\ta - 1)(\ta^* -1 )(1+ \ta + \ta^*) + 1/D\tau^v K^2}{(\ta + \ta^*)\ta \ta^*} \right] \\
    &= \left(\frac{ g_v \mu_B}{v}\right)^2 (4 \pi \sigma k_B T )\Delta_{ii'}
    \int K^2 dK e^{-2K d} 
    \left[ 1-\frac{\ta+\ta^*}{2 \ta \ta^*} \right],
\)
which goes back to Eq.~(\ref{eq:valley-noise-supp}) in the limit $1/\tau^v \rightarrow 0$. In the opposite limit $\tau^s \rightarrow 0$, the equation for the chemical potential is dominated by the intervalley fluctuation and relaxation, and essentially decouples from the intravalley fluctuations and the valley current transport. The latter continues to contribute to the magnetic noise $\widetilde{\tB}^v_{ii'} (\omega) \rightarrow (g_v \mu_B/v) (\pi \sigma k_B T/d^3) \Delta_{ii'}.$

\section{III. Simple metal}
For a paramagnetic metal, the spin fluctuations are associated with electrons on the Fermi surface and can be compared with the charge fluctuations on an equal footing. 
The spin diffusion equation remains a valid description in this scenario. We follow the discussion below Eq.~(\ref{eq:spin-noise-supp}), replacing $D^s$ and by the charge diffusion constant $D$. For a system with a long spin relaxation time, the magnetic noise induced by spin transport yields 
\(  \label{eq:ratio-spin-charge}
    \frac{B^s_{zz}}{B^c_{zz}}
    \sim \left(g_s \mu_B\right)^2 
    \frac{2 c^2 \chi_0}{e^2 \sigma \omega d^2} 
    \int d\xi \, \xi^2 e^{-2\xi} I^s(\xi, \zeta).
\)
For an estimation, we take $\chi_0$ to be the Pauli paramagnetic susceptibility $\chi_p \sim \nu$ the density of states at the Fermi surface,
\(
    \frac{B^s_{zz}}{B^c_{zz}} \sim \frac{g_s^2}{2} \left(\frac{c \alpha}{v}\right)^2
    \left(\frac{3 a_0}{\ell_\text{mfp}}\right)^2 
    \int d\xi \, \xi^2 e^{-2\xi} I^s(\xi, \zeta)/\zeta,
\)
resulting in a ratio of $0.001$ with $g_s \sim 2$, $v \sim 10^5$~m/s, $\ell_\text{mfp} \sim 30$~nm, $D\sim v \ell_\text{mfp}/3$ and $\zeta \sim 0.5$. 
The presence of strong spin-orbit interactions may lead to an extremely short spin relaxation time, such as in Pt, and thus the spin decay length can be even shorter than the electron mean free path~\cite{metal-review}. For $v\tau^s \sim \ell_\text{mfp}$, an estimation of
\(
    \frac{B^s_{zz}}{B^c_{zz}} 
    \sim (g_s \mu_B)^2 \frac{c^2 \chi_0 \tau^s}{2e^2 \sigma d^2}\;
    \overset{\chi_0 \sim \nu}{\sim} \;
    \frac{3g_s^2}{8} \left(\frac{c \alpha}{v}\right)^2
    \left(\frac{v \tau^s}{\ell_\text{mfp}}\right) 
    \left(\frac{a_0}{d}\right)^2
\)
is also typically much smaller than $1$.
Therefore, the charge fluctuations dominate the magnetic noise in most circumstances for simple metals. 
However, the spin susceptibility $\chi_0$ can be greatly enhanced near the Stoner instability, leading to a larger contribution of the spin fluctuations to the magnetic noise.

\end{document}